\documentclass[showpacs,twocolumn,preprintnumbers,amsmath,amssymb]{revtex4}

\usepackage{graphicx}
\usepackage{dcolumn}
\usepackage{bm}

\newcommand{\bq}{\begin{equation}}
\newcommand{\eq}{\end{equation}}
\newcommand{\bqa}{\begin{eqnarray}}
\newcommand{\eqa}{\end{eqnarray}}
\newcommand{\nn}{\nonumber \\}

\def\be     {\begin{equation}}
\def\ee     {\end{equation}}
\def\bea        {\begin{eqnarray}}
\def\eea        {\end{eqnarray}}
\def\bnn    {\begin{eqnarray*}}
\def\enn    {\end{eqnarray*}}

\begin{document}

\title{Stoner instability revisited: Emergence of local quantum criticality?}
\author{Ki-Seok Kim}
\affiliation{ Department of Physics, POSTECH, Pohang, Gyeongbuk 790-784, Korea \\ Institute of Edge of Theoretical Science (IES), POSTECH, Pohang, Gyeongbuk 790-784, Korea }
\date{\today}

\begin{abstract}
We revisit Stoner instability, an old problem but in a modern point of view. An idea is to extract out dynamics of directional fluctuations of spins explicitly, resorting to the CP$^{1}$ representation and integrating over their amplitude fluctuations. As a result, we derive an effective field theory for ferromagnetic quantum phase transitions in terms of bosonic spinons and fermionic holons. We show that this effective field theory reproduces overdamped spin dynamics in a paramagnetic Fermi liquid and magnon spectrum in a ferromagnetic Fermi liquid. An interesting observation is that the velocity of spinons becomes zero, approaching the ferromagnetic quantum critical point, which implies emergence of local quantum criticality. Based on this scenario, we predict the $\omega/T$ scaling behavior near ferromagnetic quantum criticality beyond the conventional scenario of the weak-coupling approach.
\end{abstract}


\maketitle

\section{Introduction}

Stoner instability is an old problem in condensed matter physics \cite{Stoner}. However, it's true that there still remain many unclarified issues in this problem. For example, the criterion for Stoner instability estimated from the mean-field analysis becomes suspected when renormalization in electron correlations is introduced \cite{Nagaosa_Book}. In addition, the first order transition takes place at low temperatures instead of the second order, which originates from fluctuation corrections \cite{BVK,Pepin_FM,Chubukov_FM}. Besides the nature of the phase transition, the dynamical critical exponent $z = 3$ quantum criticality leads the Hertz-Moriya-Millis theory to live above its upper critical dimension, giving rise to the mean-field-type behavior and violating $\omega/T$ scaling \cite{QCP_Review1,QCP_Review2}. Recently, vertex corrections have been pointed out to play an important role in ferromagnetic quantum criticality, in particular, for two dimensions, expected to modify critical exponents \cite{SungSik,Metlitski}. Frankly speaking, it's fair to say that we do not understand the nature of ferromagnetic quantum criticality and the origin of non-Fermi liquid physics near ferromagnetic quantum criticality.

In this study we revisit this old fashioned but challenging problem in a modern point of view. An idea is to extract out dynamics of directional fluctuations of spins explicitly, resorting to the CP$^{1}$ representation \cite{Auerbach} and integrating over their magnitude fluctuations. As a result, we derive an effective field theory for ferromagnetic quantum phase transitions in terms of bosonic spinons and fermionic holons. We justify this effective field theory, verifying that both overdamped spin dynamics in a paramagnetic Fermi liquid \cite{Chubukov_FL} and magnon spectrum in a ferromagnetic Fermi liquid are well reproduced. An interesting point is that the velocity of spinons vanishes, approaching the ferromagnetic quantum critical point, which implies emergence of local quantum criticality \cite{Si_LQCP}. This observation suggests an appealing scenario that directional spin fluctuations become fractionalized to allow bosonic spinon excitations and their dynamics is locally critical at the ferromagnetic quantum critical point as long as spinon excitations remain deconfined against gauge fluctuations and such gauge fluctuations do not alter the nature of the ferromagnetic transition. Based on this scenario, we predict the $\omega/T$ scaling behavior near ferromagnetic quantum criticality beyond the conventional scenario of the weak-coupling approach \cite{HMM}.

\section{U(1) slave spin-rotor theory}

\subsection{Formulation}

We start from a Hubbard-type model
\bqa && Z = \int D c_{i\sigma} D \boldsymbol{\Phi}_{i} \exp\Bigl[ - \int_{0}^{\beta} d \tau \Bigl\{ \sum_{i} c_{i\sigma}^{\dagger} (\partial_{\tau} - \mu) c_{i\sigma} \nn && - t \sum_{ij} (c_{i\sigma}^{\dagger} c_{j\sigma} + H.c.) - \sum_{i} c_{i\alpha}^{\dagger} \boldsymbol{\Phi}_{i} \cdot \boldsymbol{\sigma}_{\alpha\beta} c_{i\beta} \nn && + \frac{1}{2g} \sum_{i} \boldsymbol{\Phi}_{i}^{2} \Bigr\} \Bigr] , \eqa where a local interaction term is decomposed into charge and spin channels but only the spin channel is kept because charge fluctuations are assumed to be not critical.

The magnetization order parameter is represented as follows
\bqa && \boldsymbol{\Phi}_{i} \cdot \boldsymbol{\sigma}_{\alpha\beta} = m_{i} U_{i\alpha\gamma} \sigma^{3}_{\gamma\delta} U_{i\delta\beta}^{\dagger} ,  \eqa where $m_{i}$ is its magnitude and $\bm{U}_{i} = \left(\begin{array}{cc} z_{i\uparrow} & z_{i\downarrow}^{\dagger} \\ z_{i\downarrow} & - z_{i\uparrow}^{\dagger} \end{array} \right)$ is an SU(2) matrix field.

Inserting Eq. (2) into Eq. (1), it is natural to introduce
\bqa && c_{i\alpha} = U_{i\alpha\beta} f_{i\beta} , \eqa expected to be justified in the large$-g$ limit. Introduction of bosonic spinon $z_{i\sigma}$ and fermionic holon $f_{i\sigma}$ is our stating point. In section IV we discuss the applicability of the present decomposition more carefully.

It is straightforward to rewrite Eq. (1) in terms of spinons and holons as follows
\bqa && Z = \int D f_{i\alpha} D U_{i\alpha\beta} D m_{i} \nn && \exp\Bigl[ - \int_{0}^{\beta} d \tau \Bigl\{ \sum_{i} f_{i\alpha}^{\dagger} [(\partial_{\tau} - \mu) \delta_{\alpha\beta} - U_{i\alpha\gamma}^{\dagger} \partial_{\tau} U_{i\gamma\beta} ] f_{i\beta} \nn && - t \sum_{ij} (f_{i\alpha}^{\dagger} U_{i\alpha\gamma}^{\dagger} U_{j\gamma\beta} f_{j\beta} + H.c.) - \sum_{i} m_{i} f_{i\alpha}^{\dagger} \sigma^{3}_{\gamma\delta} f_{i\beta} \nn && + \frac{1}{2g} \sum_{i} m_{i}^{2} \Bigr\} \Bigr] , \eqa where no approximations have been used.

Correlations between spinons and holons are decomposed, resorting to the Hubbard-Stratonovich transformation. The time part is
\bqa && - f_{i\alpha}^{\dagger} U_{i\alpha\gamma}^{\dagger} \partial_{\tau} U_{i\gamma\beta} f_{i\beta} \rightarrow - x_{i} \sigma f_{i\sigma}^{\dagger} f_{i\sigma} + y_{i} z_{i\sigma}^{\dagger} \partial_{\tau} z_{i\sigma} - x_{i} y_{i} , \nonumber \eqa and the spatial part is
\bqa && - t \sum_{ij} (f_{i\alpha}^{\dagger} U_{i\alpha\gamma}^{\dagger} U_{j\gamma\beta} f_{j\beta} + H.c.) \nn && = - t \Bigl\{ f_{i+}^{\dagger} (z_{i\sigma}^{\dagger} z_{j\sigma}) f_{j+} + f_{i+}^{\dagger} (\epsilon_{\sigma\sigma'} z_{i\sigma}^{\dagger} z_{j\sigma'}^{\dagger}) f_{j-} \nn && + f_{i-}^{\dagger} (\epsilon_{\sigma\sigma'} z_{j\sigma} z_{i\sigma'}) f_{j+} + f_{i-}^{\dagger} (z_{j\sigma}^{\dagger} z_{i\sigma}) f_{j-} + H.c. \Bigr\} \nn && \rightarrow - t \Bigl\{ f_{i+}^{\dagger} \chi_{ij}^{f} f_{j+} + z_{i\sigma}^{\dagger} \chi_{ij}^{z +} z_{j\sigma} - \chi_{ij}^{f} \chi_{ij}^{z +} \nn && + f_{i-}^{\dagger} \chi_{ij}^{f *} f_{j-} + z_{j\sigma}^{\dagger} \chi_{ij}^{z -} z_{i\sigma} - \chi_{ij}^{f *} \chi_{ij}^{z -} + H.c. \Bigr\} , \nonumber \eqa where spin-singlet pairing fluctuations are assumed to be not relevant and neglected.

An effective theory reads
\bqa && Z = \int D f_{i\alpha} D z_{i\sigma} D \lambda_{i} D m_{i} D x_{i} D y_{i} D \chi_{ij}^{f} D \chi_{ij}^{z \pm} \nn && \exp\Bigl[ - \int_{0}^{\beta} d \tau \Bigl\{ \sum_{i} \Bigl( f_{i\sigma}^{\dagger} (\partial_{\tau} - \mu) f_{i\sigma} - \sigma ( m_{i} + x_{i}) f_{i\sigma}^{\dagger} f_{i\sigma} \Bigr) \nn && - t \sum_{ij} \Bigl( f_{i+}^{\dagger} \chi_{ij}^{f} f_{j+} + f_{i-}^{\dagger} \chi_{ij}^{f *} f_{j-}  + H.c. \Bigr) + \sum_{i} y_{i} z_{i\sigma}^{\dagger} \partial_{\tau} z_{i\sigma} \nn && - t \sum_{ij} \Bigl( z_{i\sigma}^{\dagger} \chi_{ij}^{z +} z_{j\sigma} + z_{j\sigma}^{\dagger} \chi_{ij}^{z -} z_{i\sigma} + H.c. \Bigr) \nn && + i \sum_{i} \lambda_{i} (|z_{i\sigma}|^{2} - 1) + \frac{1}{2g} \sum_{i} m_{i}^{2} \Bigr\} + \beta \sum_{i} x_{i} y_{i} \nn && - \beta t \sum_{ij} \Bigl( \chi_{ij}^{f} \chi_{ij}^{z +} + \chi_{ij}^{f *} \chi_{ij}^{z -} + H.c. \Bigr) \Bigr] , \eqa where $\lambda_{i}$ incorporates the unimodular constraint for the spinon field. Although spin-singlet pairing fluctuations are neglected, relevant approximations have not been made. In other words, integration for $x_{i}$, $y_{i}$, $\chi_{ij}^{f}$, and $\chi_{ij}^{z \pm}$ recovers Eq. (4) essentially.

Next, we perform integrals for $x_{i}$ and $y_{i}$. But, we determine hopping parameters of $\chi_{ij}^{f}$ and $\chi_{ij}^{z \pm}$ in the saddle-point approximation, resulting in band renormalization for spinons and holons. The saddle-point value of $i \lambda_{i} \rightarrow \lambda$ plays the role of mass in spinon excitations. Then, we reach the following expression
\bqa && Z = \int D f_{i\alpha} D z_{i\sigma} D m_{i} \nn && \exp\Bigl[ - \int_{0}^{\beta} d \tau \Bigl\{ \sum_{i} f_{i\sigma}^{\dagger} (\partial_{\tau} - \mu - \sigma m_{i}) f_{i\sigma} - t \sum_{ij} ( f_{i\sigma}^{\dagger} \chi_{ij}^{f} f_{j\sigma} \nn && + H.c ) + \frac{1}{2g} \sum_{i} \Bigl( z_{i\sigma}^{\dagger} \partial_{\tau} z_{i\sigma} - \frac{m_{i}}{2} \Bigr)^{2} - t \sum_{ij} ( z_{i\sigma}^{\dagger} \chi_{ij}^{z} z_{j\sigma} + H.c. \Bigr) \nn && + \lambda \sum_{i} |z_{i\sigma}|^{2} \Bigr\} - \beta \Bigl\{ t \sum_{ij} \Bigl( \chi_{ij}^{f} \chi_{ij}^{z} + H.c. \Bigr) - L^{d} \lambda \Bigr\} \Bigr] , \eqa where the saddle-point analysis for hopping parameters of spinons gives \bqa && \chi_{ij}^{z +} = \chi_{ij}^{z - *} = \chi_{ij}^{z} . \eqa

We call this formulation U(1) slave spin-rotor theory, the name of which is to benchmark U(1) slave-rotor theory for charge fluctuations \cite{U1SR_Florens}. Unfortunately, U(1) slave spin-rotor theory turns out to be not stable in contrast with U(1) slave charge-rotor theory. The positive sign in $\frac{1}{2g} \sum_{i} \Bigl( z_{i\sigma}^{\dagger} \partial_{\tau} z_{i\sigma} - \frac{m_{i}}{2} \Bigr)^{2}$ favors stronger directional fluctuations while it is negative in the U(1) slave-rotor theory, serving a parabolic potential for charge fluctuations. This difference originates from the opposite sign when the Hubbard-$U$ term is decomposed into charge and spin channels.

\subsection{Integration of $m_{i}$}

An important point of this study is to overcome the inconsistency of U(1) slave spin-rotor formulation, performing integrals for amplitude fluctuations.
We consider amplitude fluctuations $\delta m_{i}$ around average magnetization $m$ as follows
\bqa && Z = \int D f_{i\alpha} D z_{i\sigma} D \delta m_{i} \exp\Bigl[ - \int_{0}^{\beta} d \tau \Bigl\{ \sum_{i} f_{i\sigma}^{\dagger} (\partial_{\tau} - \mu \nn && - \sigma m - \sigma \delta m_{i}) f_{i\sigma} - t \sum_{ij} ( f_{i\sigma}^{\dagger} \chi_{ij}^{f} f_{j\sigma} + H.c ) \nn && + \frac{1}{2g} \sum_{i} \Bigl( z_{i\sigma}^{\dagger} \partial_{\tau} z_{i\sigma} - \frac{m}{2} - \frac{\delta m_{i}}{2} \Bigr)^{2} \nn && - t \sum_{ij} ( z_{i\sigma}^{\dagger} \chi_{ij}^{z} z_{j\sigma} + H.c. \Bigr) + \lambda \sum_{i} |z_{i\sigma}|^{2} \Bigr\} \nn && - \beta \Bigl\{ t \sum_{ij} \Bigl( \chi_{ij}^{f} \chi_{ij}^{z} + H.c. \Bigr) - L^{d} \lambda \Bigr\} \Bigr] . \eqa

It is straightforward to perform a self-consistent RPA (random phase approximation) analysis for $\delta m_{i}$, giving rise to the following effective field theory
\begin{widetext}
\bqa && Z = \int D f_{i\alpha} D z_{i\sigma} \exp\Bigl[ - \int_{0}^{\beta} d \tau \Bigl\{ \sum_{i} f_{i\sigma}^{\dagger} (\partial_{\tau} - \mu - \sigma m ) f_{i\sigma} - t \sum_{ij} ( f_{i\sigma}^{\dagger} \chi_{ij}^{f} f_{j\sigma} + H.c ) \nn && - \frac{1}{2} \int_{0}^{\beta} d \tau' \sum_{i} \sum_{j} [\sigma f_{i\sigma}^{\dagger} f_{i\sigma}]_{\tau} D_{ij} (\tau-\tau';m) [\sigma' f_{j\sigma}^{\dagger} f_{j\sigma}]_{\tau'} \Bigr\} - \frac{1}{2} \sum_{i\Omega} \sum_{\boldsymbol{q}} \Bigl\{ \ln \Bigl( \frac{1}{4g} - \Pi(\boldsymbol{q},i\Omega;m) \Bigr) + \Pi(\boldsymbol{q},i\Omega;m) D(\boldsymbol{q},i\Omega;m) \Bigr\} \nn && - \int_{0}^{\beta} d \tau \Bigl\{ \frac{1}{2g} \sum_{i} ( z_{i\sigma}^{\dagger} \partial_{\tau} z_{i\sigma} )^{2} - \frac{1}{8 g^{2}} \int_{0}^{\beta} d \tau' \sum_{i} \sum_{j} ( z_{i\sigma}^{\dagger} \partial_{\tau} z_{i\sigma} )_{\tau} \Bigl( \frac{1}{4g} \boldsymbol{I} - \boldsymbol{\Pi} \Bigr)^{-1}_{\tau\tau',ij} ( z_{j\sigma'}^{\dagger} \partial_{\tau'} z_{j\sigma'} )_{\tau'} \nn && - \frac{m^{2}}{32 g^{2}} \int_{0}^{\beta} d \tau' \sum_{i} \sum_{j} \Bigl( \frac{1}{4g} \boldsymbol{I} - \boldsymbol{\Pi} \Bigr)^{-1}_{\tau\tau',ij} - \frac{m}{2 g} \sum_{i} ( z_{i\sigma}^{\dagger} \partial_{\tau} z_{i\sigma} ) + \frac{m}{8g^{2}} \int_{0}^{\beta} d \tau' \sum_{i} \sum_{j} ( z_{i\sigma}^{\dagger} \partial_{\tau} z_{i\sigma} )_{\tau} \Bigl( \frac{1}{4g} \boldsymbol{I} - \boldsymbol{\Pi} \Bigr)^{-1}_{\tau\tau',ij} \nn && - t \sum_{ij} ( z_{i\sigma}^{\dagger} \chi_{ij}^{z} z_{j\sigma} + H.c. \Bigr) + \lambda \sum_{i} |z_{i\sigma}|^{2} \Bigr\} - \beta \Bigl\{ t \sum_{ij} \Bigl( \chi_{ij}^{f} \chi_{ij}^{z} + H.c. \Bigr) + L^{d} \frac{m^{2}}{8g} - L^{d} \lambda \Bigr\} \Bigr] , \eqa where all undefined symbols will be clarified below.

Resorting to the uniform ansatz of $\chi_{ij}^{f} = \chi_{f}$ and $\chi_{ij}^{z} = \chi_{z}$, and performing the Fourier transformation, we reach the following expression
\bqa && Z = \int D f_{\boldsymbol{k}\alpha} D z_{\boldsymbol{q}\sigma} \exp\Bigl[ - \sum_{i\omega} \sum_{\boldsymbol{k}} \Bigl\{ f_{\boldsymbol{k}\sigma}^{\dagger} (- i \omega - \mu - t \chi_{f} \gamma_{\boldsymbol{k}} - \sigma m) f_{\boldsymbol{k}\sigma} \nn && - \frac{1}{2} \sum_{i\Omega} \frac{1}{\beta} \sum_{i\omega} \frac{1}{\beta} \sum_{i\omega'} \sum_{\boldsymbol{q}} \sum_{\boldsymbol{k}} \sum_{\boldsymbol{k}'} [\sigma f_{\boldsymbol{k}\sigma}^{\dagger}(i\omega) f_{\boldsymbol{k}+\boldsymbol{q}\sigma}(i\omega+i\Omega)] D(\boldsymbol{q},i\Omega;m) [\sigma' f_{\boldsymbol{k}'\sigma'}^{\dagger}(i\omega') f_{\boldsymbol{k}'-\boldsymbol{q}\sigma'}(i\omega'-i\Omega)] \Bigr\} \nn && - \frac{1}{2} \sum_{i\Omega} \sum_{\boldsymbol{q}} \Bigl\{ \ln \Bigl( \frac{1}{4g} - \Pi(\boldsymbol{q},i\Omega;m) \Bigr) + \Pi(\boldsymbol{q},i\Omega;m) D(\boldsymbol{q},i\Omega;m) \Bigr\} \nn && - \sum_{i\Omega} \sum_{\boldsymbol{q}} \Bigl\{ \frac{m}{2 g} \frac{\Pi(0,0;m)}{\frac{1}{4g} - \Pi(0,0;m)} z_{\boldsymbol{q}\sigma}^{\dagger} \partial_{\tau} z_{\boldsymbol{q}\sigma} - \frac{1}{4g} ( z_{i\sigma}^{\dagger} \partial_{\tau} z_{i\sigma} )_{-\boldsymbol{q},-i\Omega} \frac{\Pi(\boldsymbol{q},i\Omega;m)}{\frac{1}{4g} - \Pi(\boldsymbol{q},i\Omega;m)} ( z_{j\sigma'}^{\dagger} \partial_{\tau'} z_{j\sigma'} )_{\boldsymbol{q},i\Omega} + ( \lambda - t \chi_{z} \gamma_{\boldsymbol{q}} ) z_{\boldsymbol{q}\sigma}^{\dagger} z_{\boldsymbol{q}\sigma} \Bigr\} \nn && - \beta L^{d} \Bigl\{ 2 z t \chi_{f} \chi_{z} - \lambda - \frac{m^{2}}{8g} \frac{\Pi(0,0;m)}{\frac{1}{4g} - \Pi(0,0;m)} \Bigr\} \Bigr] . \eqa
$t \chi_{f} \gamma_{\boldsymbol{k}}$ is the dispersion of holons and $t \chi_{z} \gamma_{\boldsymbol{q}}$ is that of spinons. $\Pi(\boldsymbol{q},i\Omega;m)$ and $D(\boldsymbol{q},i\Omega;m)$ are the self-energy and propagator for amplitude fluctuations, respectively.

An essential modification is that the positive sign in $\frac{1}{2g} \sum_{i} \Bigl( z_{i\sigma}^{\dagger} \partial_{\tau} z_{i\sigma} - \frac{m_{i}}{2} \Bigr)^{2}$ of Eq. (6) turns into the negative sign in $- \frac{1}{4g} ( z_{i\sigma}^{\dagger} \partial_{\tau} z_{i\sigma} )_{-\boldsymbol{q},-i\Omega} \frac{\Pi(\boldsymbol{q},i\Omega;m)}{\frac{1}{4g} - \Pi(\boldsymbol{q},i\Omega;m)} ( z_{j\sigma'}^{\dagger} \partial_{\tau'} z_{j\sigma'} )_{\boldsymbol{q},i\Omega}$ of Eq. (10) for the paramagnetic state $m = 0$. As a result, U(1) slave spin-rotor theory becomes consistent, i.e., stable in the description for spin dynamics.

Integrating over spinons and holons, we obtain the following expression of the Luttinger-Ward functional free energy \cite{LW_Functional}
\bqa && F(m,\lambda;g,T) = - \frac{N_{\sigma}}{2\beta} \sum_{i\omega} \sum_{\bm{k}} \Bigl[ \ln \Bigl\{ - i \omega - \mu - t \chi_{f} \gamma_{\boldsymbol{k}} - \sigma m + \Sigma_{f}(\bm{k},i\omega;m) \Bigr\} + \Sigma_{f}(\bm{k},i\omega;m) G_{f}(\bm{k},i\omega;m) \Bigr] \nn && + \frac{1}{2\beta} \sum_{i\Omega} \sum_{\boldsymbol{q}} \ln \Bigl( \frac{1}{4g} - \Pi(\boldsymbol{q},i\Omega;m) \Bigr) + \frac{N_{\sigma}}{\beta} \sum_{i\omega} \sum_{\bm{k}} \ln \Bigl\{ \frac{m}{2 g} \frac{\Pi(0,0;m)}{\frac{1}{4g} - \Pi(0,0;m)} (- i \omega) + \lambda - t \chi_{z} \gamma_{\boldsymbol{k}} + \Sigma_{z}(\bm{k},i\omega;m) \Bigr\} \nn && + L^{d} \Bigl\{ 2 z t \chi_{f} \chi_{z} - \lambda - \frac{m^{2}}{8g} \frac{\Pi(0,0;m)}{\frac{1}{4g} - \Pi(0,0;m)} \Bigr\} , \eqa
where $\Sigma_{f}(\bm{k},i\omega;m)$ and $G_{f}(\bm{k},i\omega;m)$ are the self-energy and Green's function of holons.
The amplitude-fluctuation self-energy $\Pi(\boldsymbol{q},i\Omega;m)$, the holon self-energy $\Sigma_{f\sigma}(\bm{k},i\omega;m)$, and the spinon self-energy $\Sigma_{z}(\bm{k},i\omega;m)$ are given by the self-consistent noncrossing approximation \cite{KS_Pepin}
\bqa && \Pi(\boldsymbol{q},i\Omega;m) = \frac{N_{\sigma}}{2\beta} \sum_{i\omega} \sum_{\bm{k}} G_{f\sigma}(\bm{k}+\bm{q},i\omega+i\Omega;m) G_{f\sigma}(\bm{k},i\omega;m) , \nn && \Sigma_{f\sigma}(\bm{k},i\omega;m) = - \frac{1}{\beta} \sum_{i\Omega} \sum_{\bm{q}} G_{f\sigma}(\bm{k}+\bm{q},i\omega+i\Omega;m) D(\bm{q},i\Omega;m) , \nn && \Sigma_{z}(\bm{k},i\omega;m) = - \frac{1}{\beta} \sum_{i\Omega} \sum_{\bm{q}} G_{z}(\bm{k}-\bm{q},i\omega-i\Omega;m) \frac{(i\omega+i\Omega)^{2} \Pi(\boldsymbol{q},i\Omega;m)}{1 - 4 g \Pi(\boldsymbol{q},i\Omega;m)} , \eqa where the holon Green's function $G_{f\sigma}(\bm{k},i\omega;m)$, the amplitude-fluctuation propagator $D(\bm{q},i\Omega;m)$, and the spinon Green's function $G_{z}(\bm{k},i\omega;m)$ are \bqa && G_{f\sigma}(\bm{k},i\omega;m) = \frac{1}{ i \omega + \mu + t \chi_{f} \gamma_{\boldsymbol{k}} + \sigma m - \Sigma_{f\sigma}(\bm{k},i\omega;m)} , \nn && D(\bm{q},i\Omega;m) = \frac{1}{\frac{1}{4g} - \Pi(\boldsymbol{q},i\Omega;m)} , \nn && G_{z}(\bm{k},i\omega;m) = \frac{1}{\frac{m}{2 g} \frac{\Pi(0,0;m)}{\frac{1}{4g} - \Pi(0,0;m)} (- i \Omega) + \lambda - t \chi_{z} \gamma_{\boldsymbol{k}} + \Sigma_{z}(\bm{k},i\omega;m)} . \eqa
\end{widetext}

Both contributions from holons and amplitude fluctuations correspond to those of the conventional weak-coupling approach in the self-consistent RPA framework. On the other hand, the last contribution from directional spin fluctuations is newly but naturally introduced in the U(1) slave spin-rotor representation. We discuss the spin dynamics associated with directional spin fluctuations.

\section{Fermi liquids revisited}

\subsection{Dynamics of spin fluctuations in a paramagnetic Fermi liquid}

In order to justify the validity of U(1) slave spin-rotor theory, we check out whether this effective field theory recovers the well-known overdamped spin dynamics in Fermi liquid \cite{Chubukov_FL} or not. It is natural to consider condensation of spinons in Fermi liquid and to take into account their gaussian fluctuations, identified with Goldstone modes. The spinon condensation will be verified below in the nonlinear $\sigma$-model approach. It is convenient to consider the easy-plane limit $z_{i\sigma} = e^{i \theta_{i\sigma}}$ for the description of sound-type modes associated with spin fluctuations, which will not change the dispersion of spin fluctuations in Fermi liquid. The U(1) slave spin-rotor theory for Fermi liquid is
%
%
\bqa && Z = \int D f_{\boldsymbol{k}\alpha} D \theta_{\boldsymbol{q}\sigma} \exp\Bigl[ - \sum_{i\omega} \sum_{\boldsymbol{k}} \Bigl\{ f_{\boldsymbol{k}\sigma}^{\dagger} (- i \omega - \mu \nn && - t \chi_{f} \gamma_{\boldsymbol{k}}) f_{\boldsymbol{k}\sigma} - \frac{1}{2} \sum_{i\Omega} \frac{1}{\beta} \sum_{i\omega} \frac{1}{\beta} \sum_{i\omega'} \sum_{\boldsymbol{q}} \sum_{\boldsymbol{k}} \sum_{\boldsymbol{k}'} \nn && [\sigma f_{\boldsymbol{k}\sigma}^{\dagger}(i\omega) f_{\boldsymbol{k}+\boldsymbol{q}\sigma}(i\omega+i\Omega)] D(\boldsymbol{q},i\Omega) \nn && [\sigma' f_{\boldsymbol{k}'\sigma'}^{\dagger}(i\omega') f_{\boldsymbol{k}'-\boldsymbol{q}\sigma'}(i\omega'-i\Omega)] \Bigr\} \nn && - \frac{1}{2} \sum_{i\Omega} \sum_{\boldsymbol{q}} \Bigl\{ \ln \Bigl( \frac{1}{4g} - \Pi(\boldsymbol{q},i\Omega) \Bigr) + \Pi(\boldsymbol{q},i\Omega) D(\boldsymbol{q},i\Omega) \Bigr\} \nn && - \sum_{i\Omega} \sum_{\boldsymbol{q}} \theta_{\boldsymbol{q}\sigma} \Bigl( \frac{\Omega^{2} \Pi(\boldsymbol{q},i\Omega)}{1 - 4 g \Pi(\boldsymbol{q},i\Omega)} + t \chi_{z} q^{2} \Bigr) \theta_{-\boldsymbol{q}\sigma} \nn && - \beta L^{d} 2 z t \chi_{f} \chi_{z} \Bigr] , \eqa where the mass $\lambda$ of spinons is set to lead the spinon condensation and the average magnetization is zero.

The Luttinger-Ward functional in the U(1) slave spin-rotor description for Fermi liquid is straightforward to read
\bqa && F(g,T) = - \frac{N_{\sigma}}{\beta} \sum_{i\omega} \sum_{\bm{k}} \Bigl[ \ln \Bigl\{ - i \omega - \mu - t \chi_{f} \gamma_{\boldsymbol{k}} \nn && + \Sigma_{f}(\bm{k},i\omega) \Bigr\} + \Sigma_{f}(\bm{k},i\omega) G_{f}(\bm{k},i\omega) \Bigr] \nn && + \frac{1}{2\beta} \sum_{i\Omega} \sum_{\boldsymbol{q}} \ln \Bigl( \frac{1}{4g} - \Pi(\boldsymbol{q},i\Omega) \Bigr) \nn && + \frac{N_{\sigma}}{2\beta} \sum_{i\Omega} \sum_{\bm{q}} \ln \Bigl\{ \frac{\Omega^{2} \Pi(\boldsymbol{q},i\Omega)}{1 - 4 g \Pi(\boldsymbol{q},i\Omega)} + t \chi_{z} q^{2} \Bigr\} \nn && + L^{d} 2 z t \chi_{f} \chi_{z} . \eqa As discussed before, both holon and amplitude-fluctuation contributions in the free energy correspond to those of Fermi liquid in the context of the spin-fermion model, where one-loop corrections are introduced self-consistently \cite{KS_Pepin}. The spinon free energy is now well defined, shown from $1 - 4 g \Pi(0,0) > 0$, while the coefficient of the $\Omega^{2}$ term is negative in Eq. (6), giving rise to inconsistency.

The dispersion of spin fluctuations in Fermi liquid is determined by
\bqa && \frac{\Omega^{2} \Pi(\boldsymbol{q},i\Omega)}{1 - 4 g \Pi(\boldsymbol{q},i\Omega)} + t \chi_{z} q^{2} = 0 , \eqa where the polarization function is well known to be the form of Landau damping, given by
\bqa && \Pi(\bm{q},i\Omega) = N_{\sigma} \frac{k_{F}^{f2}}{2\pi^{2}v_{F}^{f}} \Bigl\{1 - \frac{1}{2} \frac{i\Omega}{v_{F}^{f} q} \ln\Bigl( \frac{i\Omega+v_{F}^{f}q}{i\Omega-v_{F}^{f}q} \Bigr) \Bigr\} \nn && \approx N_{\sigma} \frac{k_{F}^{f2}}{2\pi^{2}v_{F}^{f}} \Bigl( 1 + \frac{\pi}{v_{F}^{f}} \frac{|\Omega|}{q} \Bigr) \eqa in three dimensions \cite{Many_Body_Textbook}. $N_{\sigma}$ represents the spin degeneracy, and $k_{F}^{f}$ and $v_{F}^{f}$ are the Fermi momentum and Fermi velocity of spinons, respectively, both of which can be expressed by $t \chi_{f}$. Then, we obtain
\bqa && |\Omega| = v_{sf} |\bm{q}| \eqa in the Matsubara frequency, implying overdamping in spin dynamics. This overdamped spin dynamics is consistent with that of the RPA theory for Fermi liquid \cite{Chubukov_FL}. Such spin dynamics turns out to originate from the Landau damping term. In other words, the well-propagating sound mode results from the spinon condensation in the absence of the Landau damping term, as expected.

The spin-fluctuation velocity is given by
\bqa && v_{sf} = \frac{1}{N_{\sigma} \frac{k_{F}^{f2}}{2\pi^{2}v_{F}^{f}}} \Bigl[ g N_{\sigma} t \chi_{z} \frac{k_{F}^{f2}}{ \pi v_{F}^{f 2}} -\Bigl\{ \Bigl( g N_{\sigma} t \chi_{z} \frac{k_{F}^{f2}}{ \pi v_{F}^{f 2}} \Bigr)^2 \nn && - N_{\sigma} t \chi_{z} \frac{k_{F}^{f2}}{2\pi^{2}v_{F}^{f}} \Bigl(1 - 4 g N_{\sigma} \frac{k_{F}^{f2}}{2\pi^{2}v_{F}^{f}} \Bigr) \Bigr\}^{1/2} \Bigr] . \eqa Our key observation is that the velocity vanishes, approaching the quantum critical point defined by $1 - 4 g \Pi(0,0) = 0$. This result leads us to suspect emergence of local quantum criticality in the problem of Stoner instability.

\subsection{Dynamics of spin fluctuations in the nonlinear $\sigma$-model approach}

Although sound-type spin fluctuations in Fermi liquid with spinon condensation are shown to reproduce the overdamped spin dynamics of Fermi liquid, the spinon condensation itself is not verified yet. In order to confirm that spinons become condensed in Fermi liquid, we benchmark the nonlinear $\sigma-$model approach of the U(1) slave charge-rotor theory \cite{U1SR_Florens}. It is straightforward to obtain the Luttinger-Ward functional within the nonlinear $\sigma-$model approach
\bqa && F(\lambda;g,T) \nn && = - \frac{N_{\sigma}}{\beta} \sum_{i\omega} \sum_{\bm{k}} \Bigl[ \ln \Bigl\{ - i \omega - \mu - t \chi_{f} \gamma_{\boldsymbol{k}} + \Sigma_{f}(\bm{k},i\omega) \Bigr\} \nn && + \Sigma_{f}(\bm{k},i\omega) G_{f}(\bm{k},i\omega) \Bigr] + \frac{1}{2\beta} \sum_{i\Omega} \sum_{\boldsymbol{q}} \ln \Bigl( \frac{1}{4g} - \Pi(\boldsymbol{q},i\Omega) \Bigr) \nn && + \frac{N_{\sigma}}{\beta} \sum_{i\omega} \sum_{\bm{k}} \ln \Bigl\{ \lambda - t \chi_{z} \gamma_{\boldsymbol{k}} + \Sigma_{z}(\bm{k},i\omega) \Bigr\} \nn && + L^{d} ( 2 z t \chi_{f} \chi_{z} - \lambda ) , \eqa where the mass of spinons is introduced to manage the spinon condensation. We recall self-consistent equations for $\Pi(\boldsymbol{q},i\Omega)$, ${\Sigma}_{f}(\boldsymbol{k},i\omega)$, and ${\Sigma}_{z}(\boldsymbol{k},i\omega)$ in Eq. (12), where each Green's function is \bqa && {G}_{f}(\boldsymbol{k},i\omega) = \frac{1}{i \omega + \mu + z t \chi_{f} \gamma_{\boldsymbol{k}} - {\Sigma}_{f}(\boldsymbol{k},i\omega)} , \nn && D(\boldsymbol{q},i\Omega) = \frac{1}{\frac{1}{4g} - \Pi(\boldsymbol{q},i\Omega)} , \nn && {G}_{z}(\bm{k},i\omega) = \frac{1}{\lambda - z \chi_{z} t \gamma_{\bm{k}} + {\Sigma}_{z}(\boldsymbol{k},i\omega)} . \eqa

Although these coupled equations seem to be complicated, one can solve them, assuming that the amplitude-fluctuation self-energy remains unaltered, compared with that of the Fermi-liquid theory. Actually, the amplitude-fluctuation self-energy is quite robust at least away from quantum criticality as long as the existence of Fermi surfaces is guaranteed. It does not change within the Eliashberg approximation even at quantum criticality \cite{Pepin_FM}. But, its possible modification has been reported near quantum criticality recently beyond the Eliashberg approximation \cite{Metlitski}. Within the Eliashberg framework, the leading contribution of the spinon self-energy is given by \bqa && \Sigma_{z}(\bm{k},i\omega) = \Sigma_{z}(i\omega) = \Sigma_{z}(0) + \mathcal{A} \xi^{4} \omega^{2} \eqa away from ferromagnetic quantum criticality, where $\mathcal{A}$ is a constant that does not change much near the critical point. Technical details for the derivation of this self-energy are shown in appendix, where we did not take into account Landau damping for the spinon spectrum of Fermi liquid in order to avoid irrelevant complications. Two important aspects are both frequency-square ($\omega^{2}$) and correlation-length ($\xi$) dependence. Recall that the frequency dependence of the self-energy is essential for the spinon condensation since only the self-energy is frequency-dependent in the spinon free energy, controlling quantum fluctuations of spinon excitations. The $\omega^{2}$ dependence allows the condensation transition for spinons at zero temperature, and the existence of the correlation length controls such a transition. Since it is large near but sufficiently away quantum criticality, quantum fluctuations of spinon excitations become suppressed, resulting in their condensation in the Fermi-liquid side. This analysis has been performed extensively in the context of the SU(2) slave-rotor theory for a spin-liquid Mott insulator to a Fermi-liquid metal transition \cite{SU2SR_Kim}, which will not be shown here. On the other hand, this frequency dependence must be modified at the critical point, where the correlation length diverges. Dynamics of spinon excitations at the critical point is beyond the scope of this study, requiring more delicate self-consistent analysis.

An unexpected conclusion can be reached when the correlation length becomes short, allowing gap formation in spinon excitations. It is not clear at all whether or not this condensation transition can be achieved far away from quantum criticality. Although we believe that spinons become condensed in the Fermi-liquid side, we do not exclude emergence of an exotic metallic state away from quantum criticality, which may be applicable to the non-Fermi liquid state of MnSi \cite{MnSi_NFL}.

\subsection{Dynamics of spin fluctuations in a ferromagnetic Fermi liquid}

The U(1) slave spin-rotor theory reproduces the spectrum of spin-wave excitations in a ferromagnetic state, given by
\bqa && F_{sw} \approx \frac{N_{\sigma}}{\beta} \sum_{i\Omega} \sum_{\bm{k}} \ln \Bigl\{ \frac{m}{2 g} \frac{\Pi(0,0;m)}{\frac{1}{4g} - \Pi(0,0;m)} (- i \Omega) + t \chi_{z} q^{2}  \Bigr\} , \nn   \eqa
where the frequency-square term is sub-leading and neglected at low energies. It is straightforward to read the spin-wave spectrum from this expression
\bqa && i \Omega = \frac{t \chi_{z}|1 - 4 g \Pi(0,0;m)|}{2 m \Pi(0,0;m)} q^{2} , \eqa consistent with that in the weak-coupling approach.
%
%

\section{Discussion and summary}

\subsection{Spirit of U(1) slave spin-rotor representation}

An essential question is how to encode ``weakly" localized nature of spin dynamics. The Hertz-Moriya-Millis theory has been regarded as the standard theoretical framework for magnetic phase transitions in itinerant electron systems \cite{HMM}. However, it is not clear how this physical picture will be modified when dynamics of electrons starts to become localized. Based on the Schwinger boson representation for spin dynamics \cite{Auerbach}, an antiferromagnetic Mott insulating state has been discussed both intensively and extensively \cite{Read_Sachdev}. U(1) slave-fermion representation is an extension of the Schwinger boson description when there exist metallic charge carriers \cite{HFQCP_Kim}. Unfortunately, this type of descriptions fail to recover a Fermi liquid phase, giving rise to an anomalous metallic state with short-ranged antiferromagnetic correlations identified with spin gap \cite{ACL_YB,NFL_KS}.

In this study we constructed essentially the same prototype of descriptions, calling it U(1) slave spin-rotor theory, where directional spin fluctuations are extracted out explicitly. An achievement beyond the previous Schwinger-boson-type descriptions is that the U(1) slave spin-rotor theory succeeds in describing physics of Fermi liquid. Spinons cannot but be condensed in the Fermi liquid region, reproducing the overdamped spin dynamics of Fermi liquid. We speculate that the U(1) slave spin-rotor theory connects the spin dynamics of itinerant electrons with that of localized electrons. Actually, we could observe that the spinon velocity vanishes, approaching the quantum critical point, which implies emergence of local quantum criticality.

\subsection{U(1) slave spin-rotor mean-field theory: First order vs. second order}

Although the nature of the ferromagnetic transition is not an issue in this paper, we touch this important problem slightly. A mean-field free energy can be constructed under lots of assumptions as follows
\bqa && F(m;g,T) \nn && \approx - \frac{N_{\sigma}}{2\beta} \sum_{\bm{k}} \Bigl[ \ln \Bigl\{ 1 + \exp\Bigl( - \beta [- \mu - t \chi_{f} \gamma_{\boldsymbol{k}} - m] \Bigr) \Bigr\} \nn && + \ln \Bigl\{ 1 + \exp\Bigl( - \beta [- \mu - t \chi_{f} \gamma_{\boldsymbol{k}} + m] \Bigr) \Bigr\} \Bigr] \nn && + \frac{1}{2\beta} \sum_{i\Omega} \sum_{\boldsymbol{q}} \ln \Bigl( \frac{1}{4g} - \Pi(\boldsymbol{q},i\Omega;m) \Bigr) \nn && + L^{d} \Bigl\{ 2 z t \chi_{f} \chi_{z} - \frac{m^{2}}{8g} \frac{\Pi(0,0;m)}{\frac{1}{4g} - \Pi(0,0;m)} \Bigr\} , \eqa where the polarization function is given by
\bqa && \Pi(\boldsymbol{q},i\Omega;m) \nn && = \frac{N_{\sigma}}{2\beta} \sum_{i\omega} \sum_{\bm{k}} g_{f\sigma}(\bm{k}+\bm{q},i\omega+i\Omega;m) g_{f\sigma}(\bm{k},i\omega;m) \nonumber \eqa with \bqa && g_{f\sigma}(\bm{k},i\omega;m) = \frac{1}{ i \omega + \mu + t \chi_{f} \gamma_{\boldsymbol{k}} + \sigma m} . \nonumber \eqa The holon self-energy is neglected when the polarization function is evaluated. Actually, two contributions from holons and amplitude fluctuations consist of the mean-field $+$ RPA framework, expected to improve the saddle-point analysis \cite{HF_Multiple_Scales_Kim}.
%
%
Sound-type (Goldstone) fluctuations are neglected in the lowest-order approximation, but introduction of such fluctuations gives rise to reduction of the average magnetization.

Although this mean-field $+$ RPA theory looks simple, corrections from amplitude fluctuations introduce vertex corrections self-consistently. Actually, this issue has been discussed intensively, showing that the second order turns into the first order generically for ferromagnetic quantum phase transitions \cite{BVK}. However, we would like to emphasize that there exists another conclusion, pointing out incompleteness of the spin-fermion model since it does not satisfy the spin conservation, where spin nematicity may appear before the ferromagnetic phase transition \cite{Chubukov_FM}, regarded to be another source resulting in complications for the nature of ferromagnetic quantum phase transitions. In addition, spin-wave fluctuations must be incorporated for the phase transition in the context of the U(1) slave spin-rotor theory. We believe that this analysis should be performed sincerely near future.

\subsection{Emergence of local quantum criticality at the ferromagnetic transition?}

Spinon condensation is shown to occur in both paramagnetic Fermi liquid and ferromagnetic Fermi liquid.
%
%
In addition, the overdamped spin dynamics of Fermi liquid is recovered and the dispersion of spin-wave excitations is consistent with that of ferromagnons in the conventional approach without the U(1) slave spin-rotor representation. On the other hand, the velocity of sound-type spin fluctuations turns out to vanish, which implies that spin fluctuations become localized, approaching the quantum critical point from the Fermi liquid side. This locality issue has not been verified from the ferromagnetic side, where both the frequency-linear term and the spinon self-energy should be taken into account on equal footing near the quantum critical point.

This discussion leads us to propose an interesting scenario for ferromagnetic quantum phase transitions. We emphasize that U(1) slave spin-rotor theory should reproduce physics of both paramagnetic and ferromagnetic Fermi liquids because the spinon condensation occurs, where the description based on Goldstone fluctuations in each phase can be regarded as just another way in describing spin fluctuations of both phases. However, dynamics of spin fluctuations would be different at the quantum critical point described by U(1) slave spin-rotor theory. Spin fluctuations are assumed to be fractionalized into spinon excitations \cite{DQCP}. Surprisingly, the velocity of spinons seems to vanish approaching the quantum critical point. The present study suggests emergence of deconfined local quantum criticality \cite{DLQCP_Kim} at the ferromagnetic quantum critical point if the nature of the ferromagnetic transition remains to be the second order. Of course, we did not evaluate the spinon self-energy at the quantum critical point, which requires self-consistency. This direction of future researches will shed light on novel emergent physics in ferromagnetic transitions.

We would like to emphasize that the emergence of local quantum criticality is not accidental but ``inevitable" strongly speaking, deeply related with the consistency of the U(1) slave spin-rotor theory. The introduction of quantum corrections into the spinon dynamics is essential for the consistency of the U(1) slave spin-rotor theory, which originates from amplitude fluctuations of the ferromagnetic order parameter. Recalling the fact that the ferromagnetic quantum phase transition is described by the RPA propagator of amplitude fluctuations, the renormalized dynamics of spinons via amplitude fluctuations encodes the information of the amplitude-fluctuation RPA propagator in the time part of the consistent U(1) slave spin-rotor theory and the signature of the ferromagnetic quantum phase transition in the RPA propagator results in local quantum criticality. If we argue this connection more strongly, one may say that the consistency of the U(1) slave spin-rotor theory results in deconfined local quantum criticality in the ferromagnetic quantum phase transition.

The origin of local quantum criticality may be traced in a recent study on heavy-fermion quantum criticality \cite{HF_LQCP}. Based on the nonlinear $\sigma-$model description for dynamics of localized spins, Kondo fluctuations are shown to cause nonlocal interactions between spin fluctuations. Performing the renormalization group analysis, nonlocal interactions between spin-wave modes turn out to make the spin-wave velocity vanish logarithmically at the quantum critical point. Emergence of nonlocal interactions between spin fluctuations is not limited on the Kondo-lattice problem. It turns out that Fermi-surface fluctuations generate nonlocal interactions between order-parameter fluctuations in Hertz-Moriya-Millis theory \cite{BVK,Pepin_FM,Chubukov_FM}. Such nonlocal interactions may cause local quantum criticality. Intuitively speaking, nonlocal effective interactions are expected to enhance the coordination number, justifying the mean-field picture but dynamical because such long-ranged interactions appear at finite frequencies. This reminds us of the fact that dynamical mean-field theory becomes ``exact" at infinite dimensions \cite{DMFT_Review}.

Another important issue is on the role of gauge fluctuations in local quantum criticality. Two possibilities can be considered. If the local quantum criticality is preserved, gauge fluctuations will not play any roles, where the kinetic-energy term of spinons vanishes identically. On the other hand, they may become relevant approaching the quantum critical point from the Higgs phase. In this case, we must take into account gauge fluctuations in the Higgs phase, where the condensation amplitude of Higgs bosons (spinons) would be small. One possibility is that the second-order nature turns into the first order, known to be the fluctuation-induced first-order transition \cite{Fluctuation_First_Order} or Coleman-Weinberg mechanism \cite{CM_Mechanism}.

\subsection{Prediction}

Although we did not ``derive" a local critical field theory in terms of spinons near ferromagnetic quantum criticality, we speculate the appearance of the $\omega/T$ scaling behavior in optical conductivity and dynamic spin susceptibility. This expectation is based on the fact that the quantum critical point is not a mean-field type but an interacting fixed point due to the local quantum criticality. We admit that this should be verified more carefully. More interestingly, an essential prediction is that critical exponents would be enhanced due to the appearance of deconfined spinons since correlation functions in terms of such deconfined degrees of freedom are expressed by combinations of multi-particle propagators \cite{DQCP}. Critical exponents of the optical conductivity and the dynamical spin susceptibility can be larger than expected.

\subsection{Criticism}

One cautious person may criticize the usefulness of deconfined local quantum criticality for ferromagnetic quantum phase transitions since the nature of them is expected to be the first order. Although we agree with the fact that the deconfined quantum critical point may not emerge at zero temperature, we speculate that its trace can be found at finite temperatures in the vicinity of ferromagnetic quantum phase transitions. We expect that the $\omega/T$ scaling behavior with enhanced critical exponents appears in a certain range of temperatures. This conjecture reminds us of a recent study, which demonstrates a critical behavior in electrical resistivity near Mott metal-insulator transition, in particular, at finite temperatures \cite{LQCP_MIT}. Although the Mott metal-insulator transition is the first order at zero temperature in dynamical mean-field theory, such a machinery verified the emergence of unexpected finite-temperature critical transport for a certain temperature domain near the Mott metal-insulator phase boundary. Of course, they are completely unclear possible similarities between ferromagnetic quantum phase transitions and Mott metal-insulator transitions. However, we cannot exclude possible mathematical similarities between them, described by U(1) slave spin-rotor formulation and U(1) slave charge-rotor theory, respectively.

\subsection{Summary}

We revisited the Hertz-Moriya-Millis theoretical framework for Stoner instability and reformulated this effective field theory in terms of deconfined degrees of freedom such as fermionic holons and bosonic spinons. This U(1) slave spin-rotor theory turns out to reproduce not only the spin-wave spectrum of the ferromagnetic state but also the overdamped spin dynamics of Fermi liquid, where the spinon condensation is achieved. This successful description for Fermi liquid physics can be regarded as an improvement beyond the similar theoretical framework of the slave-fermion representation. A key observation was that the velocity of spin fluctuations vanishes approaching the quantum critical point. This leads us to propose an exotic scenario of deconfined local quantum criticality, where localized deconfined spinon excitations appear. Recall that spin$-1$ overdamped fluctuations in Fermi liquid and spin$-1$ well-propagating ferromagnon excitations in the ferromagnetic state are allowed, both of which are conventional degrees of freedom. We speculated that the origin of this deconfined local quantum criticality is the existence of nonlocal effective interactions between spin-fluctuation modes in the Hertz-Moriya-Millis theoretical framework, which arise from vertex corrections of Fermi-surface fluctuations beyond the conventional Hertz-Moriya-Millis theory. Our suggestion of deconfined local quantum criticality implies the existence of the $\omega/T$ scaling physics with enhanced critical exponents near ferromagnetic quantum criticality beyond the conventional scenario based on the weak-coupling approach. Spin dynamics at and near this ferromagnetic quantum criticality remains as the next direction of research based on the U(1) slave spin-rotor theoretical framework.

\section*{Acknowledgement}

This study was supported by the Ministry of Education, Science, and Technology (No. 2012R1A1B3000550) of the National Research Foundation of Korea (NRF) and by TJ Park Science Fellowship of the POSCO TJ Park Foundation. We appreciate fruitful collaboration with Jung-Hoon Han at the initial stage and helpful comments of D. Belitz. We also appreciate hospitality of APCTP.

\appendix*

\section{Spinon self-energy in Fermi liquid}

Inserting the polarization function of the Landau-damping form into the spinon self-energy, we obtain
\bqa && \Sigma_{z}(\bm{k},i\omega) \nn && = - N_{\sigma} \frac{k_{F}^{f2}}{2\pi^{2}v_{F}^{f}} \frac{1}{\beta} \sum_{i\Omega} \sum_{\bm{q}} \frac{1}{ \lambda - t \chi_{z} \gamma_{\boldsymbol{k}-\bm{q}} + \Sigma_{z}(\bm{k}-\bm{q},i\omega-i\Omega)} \nn && \frac{(i\omega+i\Omega)^{2}}{\Bigl( 1 - 4 g N_{\sigma} \frac{k_{F}^{f2}}{2\pi^{2}v_{F}^{f}} \Bigr) - 4 g N_{\sigma} \frac{k_{F}^{f2}}{2\pi^{2}v_{F}^{f}} \frac{\pi}{v_{F}^{f}} \frac{|\Omega|}{q} } \nn && - N_{\sigma} \frac{k_{F}^{f2}}{2\pi^{2}v_{F}^{f}} \frac{\pi}{v_{F}^{f}} \frac{1}{\beta} \sum_{i\Omega} \sum_{\bm{q}} \frac{1}{ \lambda - t \chi_{z} \gamma_{\boldsymbol{k}-\bm{q}} + \Sigma_{z}(\bm{k}-\bm{q},i\omega-i\Omega)} \nn && \frac{(i\omega+i\Omega)^{2}}{\Bigl( 1 - 4 g N_{\sigma} \frac{k_{F}^{f2}}{2\pi^{2}v_{F}^{f}} \Bigr) - 4 g N_{\sigma} \frac{k_{F}^{f2}}{2\pi^{2}v_{F}^{f}} \frac{\pi}{v_{F}^{f}} \frac{|\Omega|}{q} } \frac{|\Omega|}{q} . \eqa

This expression can be approximated in Fermi liquid as follows
\bqa && \Sigma_{z}(\bm{k},i\omega) \approx - N_{\sigma} \frac{k_{F}^{f2}}{2\pi^{2}v_{F}^{f}} \frac{1}{ 1 - 4 g N_{\sigma} \frac{k_{F}^{f2}}{2\pi^{2}v_{F}^{f}} } \nn && \frac{1}{\beta} \sum_{i\Omega} \sum_{\bm{q}} \frac{(i\omega+i\Omega)^{2}}{ \lambda - t \chi_{z} \gamma_{\boldsymbol{k}-\bm{q}} + \Sigma_{z}(\bm{k}-\bm{q},i\omega-i\Omega)} \nn && - N_{\sigma} \frac{k_{F}^{f2}}{2\pi^{2}v_{F}^{f}} \frac{\pi}{v_{F}^{f}} \frac{1}{ 1 - 4 g N_{\sigma} \frac{k_{F}^{f2}}{2\pi^{2}v_{F}^{f}} } \nn && \frac{1}{\beta} \sum_{i\Omega} \sum_{\bm{q}} \frac{(i\omega+i\Omega)^{2}}{ \lambda - t \chi_{z} \gamma_{\boldsymbol{k}-\bm{q}} + \Sigma_{z}(\bm{k}-\bm{q},i\omega-i\Omega)} \frac{|\Omega|}{q} \nn && \approx - N_{\sigma} \frac{k_{F}^{f2}}{2\pi^{2}v_{F}^{f}} \frac{1}{ 1 - 4 g N_{\sigma} \frac{k_{F}^{f2}}{2\pi^{2}v_{F}^{f}} } \nn && \frac{1}{\beta} \sum_{i\omega'} \sum_{\bm{k}'} \frac{(i\omega'+2i\omega)^{2}}{ \lambda - t \chi_{z} \gamma_{-\boldsymbol{k}'} + \Sigma_{z}(-\bm{k}',-i\omega')} , \eqa where $\frac{|\Omega|}{v_{F}^{f} q} \ll 1$ was used.

This integral equation can be solved within the ansatz of $\Sigma_{z}(\bm{k},i\omega) = \Sigma_{z}(i\omega) = \Sigma_{z}(0) + \mathcal{C} \xi^{2} \omega^{2}$. Inserting this ansatz into the above expression, we obtain
\bqa && \Sigma_{z}(0) = - N_{\sigma} \frac{k_{F}^{f2}}{2\pi^{2}v_{F}^{f}} \frac{1}{ 1 - 4 g N_{\sigma} \frac{k_{F}^{f2}}{2\pi^{2}v_{F}^{f}} } \nn && \Bigl( \frac{1}{\beta} \sum_{i\omega'} \sum_{\bm{k}'} \frac{(i\omega')^{2}}{ \lambda - t \chi_{z} \gamma_{-\boldsymbol{k}'} + \Sigma_{z}(-i\omega')} \Bigr) \eqa and
\bqa && \mathcal{C} = \frac{1}{\beta} \sum_{i\omega} \sum_{\bm{k}} \frac{1}{ \lambda - t \chi_{z} \gamma_{\boldsymbol{k}} + \Sigma_{z}(0) + \mathcal{C} \xi^{2} \omega^{2}} . \eqa

Performing the frequency summation, it is straightforward to obtain
\bqa && \mathcal{C}^{2} \xi^{2} = \sum_{\bm{k}} \frac{1}{2E_{\bm{k}}^{b}} \coth\Bigl( \frac{E_{\bm{k}}^{b}}{2T} \Bigr) . \eqa The spinon spectrum may be simplified as follows
\bqa && E_{\bm{k}}^{b} = \mathcal{C}^{-1/2} \xi^{-1} \sqrt{\lambda + \Sigma_{z}(0) - t \chi_{z} \gamma_{\boldsymbol{k}}} \nn && \equiv \mathcal{C}^{-1/2} \xi^{-1} \sqrt{m_{z}^{2} + v_{z}^{2} \bm{k}^{2}} , \eqa where $\xi^{2} \propto \Bigl(1 - 4 g N_{\sigma} \frac{k_{F}^{f2}}{2\pi^{2}v_{F}^{f}}\Bigr)^{-1}$ is the correlation length of amplitude fluctuations.

It is straightforward to perform the momentum integral
\bqa && \mathcal{C}^{2} \xi^{2} = 4 \pi \int_{0}^{\infty} d k k^{2} \frac{\mathcal{C}^{1/2} \xi}{2 \sqrt{m_{z}^{2} + v_{z}^{2} \bm{k}^{2}} } \nn && \coth\Bigl( \frac{\mathcal{C}^{-1/2} \xi^{-1} \sqrt{m_{z}^{2} + v_{z}^{2} \bm{k}^{2}}}{2 T} \Bigr) \nn && \approx 4 \pi \mathcal{C}^{3/2} \xi^{3} \int_{0}^{\Lambda} d k k^{2} \frac{1}{2 v_{z} k } \coth\Bigl( \frac{v_{z} k}{2 T} \Bigr) , \eqa which gives rise to
\bqa && \mathcal{C} \propto \xi^{2} . \eqa

\end{document}